\documentclass[twocolumn,aps,floats,superscriptaddress]{revtex4}
\usepackage{graphicx}
\usepackage{epsfig}
\usepackage{epsf}
\usepackage{amsfonts}
\usepackage{amsthm}
\usepackage{amsmath}
\usepackage{amssymb}
\usepackage[latin1]{inputenc}
\usepackage[english]{babel}

\begin{document}



\def\beq{\begin{equation}}
\def\eeq{\end{equation}}
\def\be{\begin{equation}}
\def\ee{\end{equation}}

\def\iomn{i\omega_n}
\def\iom#1{i\omega_{#1}}
\def\c#1#2#3{#1_{#2 #3}}
\def\cdag#1#2#3{#1_{#2 #3}^{+}}
\def\epsk{\epsilon_{{\bf k}}}
\def\Ga{\Gamma_{\alpha}}
\def\Seff{S_{eff}}
\def\dinf{$d\rightarrow\infty\,$}
\def\T{\mbox{Tr}}
\def\t{\mbox{tr}}
\def\cG0{{\cal G}_0}
\def\cS{{\cal S}}
\def\divnum{\frac{1}{N_s}}
\def\vac{|\mbox{vac}\rangle}
\def\intR{\int_{-\infty}^{+\infty}}
\def\intb{\int_{0}^{\beta}}
\def\spinup{\uparrow}
\def\spindown{\downarrow}
\def\bra{\langle}
\def\ket{\rangle}

\def\ka{{\bf k}}
\def\vk{{\bf k}}
\def\vq{{\bf q}}
\def\vQ{{\bf Q}}
\def\vr{{\bf r}}
\def\q{{\bf q}}
\def\R{{\bf R}}
\def\kp{\bbox{k'}}
\def\a{\alpha}
\def\b{\beta}
\def\d{\delta}
\def\D{\Delta}
\def\e{\varepsilon}
\def\eps{\epsilon}
\def\ed{\epsilon_d}
\def\ef{\epsilon_f}
\def\g{\gamma}
\def\G{\Gamma}
\def\l{\lambda}
\def\L{\Lambda}
\def\o{\omega}
\def\ph{\varphi}
\def\s{\sigma}
\def\chib{\overline{\chi}}
\def\et{\widetilde{\epsilon}}
\def\hn{\hat{n}}
\def\hnu{\hat{n}_\uparrow}
\def\hnd{\hat{n}_\downarrow}

\def\hc{\mbox{h.c}}
\def\Im{\mbox{Im}}

\def\est{\varepsilon_F^*}
\def\v2o3{V$_2$O$_3$}
\def\uc2{$U_{c2}$}
\def\uc1{$U_{c1}$}


\def\bea{\begin{eqnarray}}
\def\eea{\end{eqnarray}}
\def \bal{\begin{align}}
\def \eal{\end{align}} 
\def\#{\!\!}
\def\@{\!\!\!\!}

\def\vi{{\bf i}}
\def\vj{{\bf j}}

\def\+{\dagger}


\def\up{\spinup}
\def\down{\spindown}


\def\'{\prime}
\def\"{{\prime\prime}}


\title{Sum-rules for Raman scattering off strongly correlated electron systems}
\author{Luca de' Medici}
\affiliation{Department of Physics and Center for Materials Theory, Rutgers University, Piscataway, NJ 08854, USA}
\author{Antoine Georges}
\affiliation{Centre de Physique Théorique, Ecole Polytechnique, 91128 Palaiseau Cedex, France}
\author{Gabriel Kotliar}
\affiliation{Department of Physics and Center for Material Theory, Rutgers University, Piscataway, NJ 08854, USA}

\begin{abstract}
\vspace{0.3cm}
We investigate sum-rules applying to the Raman intensity in a strongly correlated
system close to the Mott transition. Quite generally, it can be shown that,
provided the frequency integration is performed up to a cutoff smaller than the
upper Hubbard band, a sum-rule applies to the non-resonant Raman signal of a doped Mott insulator,
resulting in an integrated intensity which is proportional to the doping level. We provide a detailed
derivation of this sum-rule for the t-J model, for which the frequency cutoff can be
taken to infinity and an unrestricted sum-rule applies.
A quantitative analysis of the sum-rule is also presented for the d-wave superconducting phase of the
t-J model, using slave boson methods.
The case of the Hubbard model is studied in the framework of dynamical
mean-field theory, with special attention to the cutoff dependence of the restricted sum-rule, and also
to the intermediate coupling regime.
The sum-rule investigated here is shown to be consistent with recent experimental data on cuprate
superconductors, reporting measurements of Raman scattering intensities on an absolute scale.

\end{abstract}

\maketitle


\vspace{2cm}

\section{Introduction}

Restricted sum-rules (relating the partial integral of the intensity up to a cutoff
with a correlation function) are useful tools for analyzing spectroscopic information
in strongly correlated materials.
For example the sum-rule for the optical conductivity has received a lot of
attention in the context of cuprates\cite{Kubo_sumrule,Norman_Pepin_sumrule,Millis_optics_lectures,VanDerMarel_optics_lectures,Deutscher_KinEn}.

In contrast the sum-rules for the Raman scattering intensity have been less studied.
These were first considered in  \cite{Kosztin_Zawadowski}\cite{Freericks_sumrules}.
Recently it has been stressed that, for a doped Mott insulator,
the non-resonant Raman intensity should be proportional to doping\cite{Shastry_Shraiman}\cite{LeTaconNaturePhysics},
provided the frequency integration is carried up to a cutoff which is such that contributions from
the upper Hubbard band are not included.
This proved instrumental in analyzing recent experiments on cuprate superconductors\cite{LeTaconNaturePhysics}, that revived the debate about the relationship between the pseudo-gap and the superconductive gap in the underdoped phase of these compounds\cite{Millis_gaps,Guyard_antinodes,Chubukov_Norman_RamanOneGap,Guyard_nodes}.

The goal of this paper is to analyze the Raman sum-rules and their dependence on the strength of the interactions, doping, temperature and choice of upper cutoff.

In section \ref{sec:proptodoping} we show that for the t-J model
the right-hand side that enters the sum-rule for Raman scattering is proportional to doping.
In this case, the frequency cutoff can be taken to infinity (since the upper Hubbard band is absent due
to the constraint of no double-occupancy), hence making a rigorous theoretical analysis easier.

We then evaluate explicitly the Raman response function of a doped Mott insulator by solving the Hubbard model using Dynamical mean-field theory.
We explore the region in which the Raman intensity scales with doping and how this region varies with the cutoff used in the sum-rule, and contrast those results with that of a correlated material slightly below the Mott transition.
In section \ref{sec:exp} we compare the results of our calculation, for different choices of the upper cutoff, with experimental data and then in section \ref{sec:KLSB} we conclude with predictions for the temperature dependence of the integrated low-energy Raman intensity, using a slave-bosons treatment of the t-J model.

\section{Raman sum-rules}

Raman scattering, is a photon-in photon-out process happening when an external electromagnetic field is applied on an system, and its non-resonant cross-section can be calculated from the Fermi golden rule, reading:
\bal
R(\vq,\Omega)&=2\pi\sum_{i,f} \frac{\exp(-\b \e_i)}{\cal{Z}} \\ \nonumber
&\times |g(\vk_i)g(\vk_f)\sum_{rs}e_r^i e_s^f\langle f|\tau^{rs}(\vq)|i\rangle|^2\d(\e_f-\e_i-\Omega),
\end{align}
where $\beta$ is the inverse temperature, $g(\vq)=(hc^2/V\o_q)^{1/2}$ where V is the volume,
${\bf \o}_q$ and ${\bf \hat e}$ are energy and polarization vectors of the photons
($i,f$ indicate initial and final states of the process) and
\be\label{eq:stress}
\tau^{rs}(\vq)=\sum_\vk\frac{\partial^2 \epsk}{\partial k_r \partial k_s}c^\+_{\vk+\vq/2, \s}c_{\vk-\vq/2, \s},
\ee
is the stress operator tensor.
$\e_{i,f}$ is the energy of the initial or final state of the system, $\Omega$ is the transferred energy,
$c^\+_\vk$ creates an electron of momentum $\vk$, $\epsk$ is the one-electron
dispersion of the model under consideration and $r,s$ are cartesian components.

Then, by the fluctuation-dissipation theorem the scattering intensity in one channel
\bal
I_{rs}(\vq,\Omega)&=\sum_{i,f} \frac{\exp(-\b \e_i)}{\cal{Z}} |\langle f|\tau^{rs}(\vq)|i\rangle|^2\d(\e_f-\e_i-\Omega)\nonumber\\
&=\frac{\chi^\"_{rs}(\vq,\Omega)}{1-e^{-\b\Omega}},
\end{align}
 is related to  the imaginary part of a response function, i.e.
\be\label{Kubo_susc}
\chi_{rs}(\vq,t)=i\Theta(t)\langle [\tau^{rs}(\vq,t),\tau^{rs}(0,0)]\rangle,
\ee
 which is a stress-stress correlation function of the unperturbed system.

We will consider in the following only the $\vq=0$ contribution, since the photon momentum is always much smaller than the Fermi momentum.

What is remarkable in Raman scattering\cite{Devereaux_RMP} is that by tuning the polarization of the incident and of the detected outgoing photons one can exploit selection rules to sort out different processes in the probed material. This is of particular utility in order to separate the response due to electronic excitations in different areas of the Brillouin zone, thus allowing to probe k-dependent properties of the material.

This is, in particular, a key issue issue for the physics of  cuprates, in which nodal and
antinodal regions of the Brillouin zone are known to behave in distinctly different manners.
It is remarkable that, despite being a $q\simeq 0$ probe (as is optics), Raman scattering
can still address momentum-selective issues by exploiting the dependence of the Raman vertex on
the polarization of the incident and scattered light.
Especially for a material with perfect in-plane equivalence of the a- and b-axis,
the $B_{2g}$ and $B_{1g}$ geometry mainly probe, respectively, the nodal and antinodal regions.

In the non-resonant response, by choosing the direction of the polarizations $e^i, e^f$,  one isolates contributions by the different elements $\tau^{rs}(\vq)$ of the stress tensor.

As an example let's consider a square lattice with nearest neighbor and next-nearest neighbor hoppings $t$ and $t^\'$, that give rise to the dispersion in k-space
\be\label{eq:disp}
\epsk=-2t(\cos{k_x}+\cos{k_y})+4t^\'\cos{k_x}\cos{k_y}.
\ee
Then in the $B_{2g}$ scattering geometry  $e^i, e^f$ are perpendicular between each other and parallel to the x-y plane axes,  and the Raman scattering operator is $\tau^{xy}(\vq)$.
From (\ref{eq:stress}) one then sees that
the sum over the momentum space is weighted by the vertex factor
\be\label{eq:B2gVertex}
\frac{\partial^2 \epsk}{\partial k_x \partial k_y}=4t^\'\sin{k_x}\sin{k_y},
\ee
 thus sorting out the contribution along the nodal directions.

Analogously one can show how in the $B_{1g}$ geometry, with  $e^i, e^f$ perpendicular, but oriented along the lattice diagonals,  one selects the antinodal directions in k-space.

In this work we are interested in discussing the sum-rules fulfilled by the Raman $B_{2g}$ non-resonant response.
It is well known\cite{Mahan, Freericks_sumrules}, that any susceptibility of the general Kubo form (\ref{Kubo_susc}) fulfills the following sum-rule:
\be
\frac{2}{\pi}\int_0^\infty d \Omega \; \Omega \; \chi^\"_{(\mathcal O,O)}(\Omega)=
\langle[{\mathcal O},[H,{\mathcal O}]]\rangle,
\ee
where $\mathcal O$ is the operator to the fluctuations of which the susceptibility is associated.

Hence, for the $B_{2g}$ raman scattering this sum-rule reads:

\be\label{eq:sumrule_omega}
\frac{2}{\pi}\int_0^\infty d \Omega \; \Omega \; \chi_{B_{2g}}^\"(\Omega)=\langle \left [ [\tau^{xy}, H], \tau^{xy} \right ]\rangle.
\ee

Eq. (\ref{eq:sumrule_omega}) has been used in Ref. \cite{LeTaconNaturePhysics} to normalize Raman spectra in mercury-based cuprates. In the following we will show that the right hand side of this sum-rule is proportional to doping in the t-t'-J model
\footnote{Using the spectral representation, one can also show that the relation
$\frac{2}{\pi}\int_0^\infty d \Omega \; \; \frac{\chi^\"_{(\mathcal O,O)}(\Omega)}{\Omega }=\chi^\'_{(\mathcal OO)}(0)$
holds as a sum-rule for any susceptibility, that is also trivially an outcome of the Kramers-Kronig relations.
Hence another sum-rule holds for the non-resonant Raman response:
$\frac{2}{\pi}\int_0^\infty d \Omega \; \frac{\chi_{B_{2g}}^\"(\Omega)}{\Omega}=\chi_{\tau^{xy}\tau^{xy}}^\'(0)$
}.

\section{The $B_{2g}$ Raman sum-rule in the $t-t^\'-J$ model and its proportionality to doping}\label{sec:proptodoping}

In this section we report the analysis of the direct evaluation of  the r.h.s. of the $B_{2g}$ sum-rule eq. (\ref{eq:sumrule_omega}) in the t-t'-J model on the square lattice.

This model is represented by the hamiltonian
\be\label{eq:tb_Ham}
H=\sum_{i,j,\s}(t_{ij} c^\+_{i,\s}c_{j,\s} + H.c.) + \sum_{kl} 4J_{kl}(S_k\cdot S_l-\frac{1}{4}n_k n_l),
\ee
acting on the Hilbert space where states with doubly occupied sites have been removed.

Here $c^\+_{i,\s}$ is the operator that creates an electron with spin $\s$ in a state localized on site $i$, and $t_{ij}$
the hopping integral between single-particle states at sites i and j,
$S_k$  is the electron spin at site k and $4J_{kl}\geq0$ is
the antiferromagnetic superexchange and is zero for $k=l$ and nonzero at (but not necessarily restricted to) nearest neighbours.

The hopping matrix $t_{ij}$ is chosen to have nonzero elements only for nearest ($t_{ij}=-t$) and next-nearest ($t_{ij}=t^\'$) neighbors. The onsite element is the chemical potential term that tunes the filling ($t_{ii}=-\mu$).  The bare electronic dispersion is given by eq. (\ref{eq:disp}).

 We have to evaluate the doping dependence of the average of the double commutator, eq. (\ref{eq:sumrule_omega}), in this model.

To implement the non-double-occupancy constraint we use the formalism of the Hubbard Operators.
Definitions and some algebra tricks for this formalism can be found in appendix \ref{app:Hubbard}. We thus obtain an exact expression of the r.h.s. of the sum-rule.

Both the stress operator (\ref{eq:stress}) and the Hamiltonian $H=H_t+H_J$ can be readily expressed with these operators $X_{\a\b}$:
 \bal
H_t &= \sum_{ij\s}t_{ij} X^i_{\s 0}X^j_{0\s}\\
H_J &= \sum_{kl, \a\b}2J_{kl} (X^k_{\a\b}X^l_{\b\a}-X^k_{\a\a}X^l_{\b\b})\\
\tau^{xy}(0)&= \sum_{ij\s}\tau^{xy}_{ij} X^i_{\s 0}X^j_{0\s}
\end{align}
where the stress component $\tau^{xy}$ has elements
\be
\tau^{xy}_{ij}=\sum_k e^{i\vk\cdot \vr_{ij}}\frac{\partial^2 \epsk}{\partial k_x \partial k_y}=4t^\'\sum_\vk \cos{(\vk\cdot \vr_{ij})}\sin{k_x}\sin{k_y}.
\ee

that are nonzero only when i and j are next-nearest neighbour, as can be deduced by this expression (in order to lighten the notation we will drop the superscript $xy$ in $\tau^{xy}_{ij}$ from now on).

The evaluation of the commutators leads, for the hopping part\footnote{Note that the diagonal terms of the Hamiltonian - e.g. the  chemical potential term - commute with the stress operator, hence they do not contribute to the sum rule. We will consider $t_{ii}=0$ from now on.}:
\begin{widetext}
\begin{multline}\label{eq:HtTau}
\left [ [\tau^{xy}, H_t], \tau^{xy} ]\right]=
\sum_{abcd,\a\neq\b}\@[(\tau_{ab}t_{bc}-\tau_{bc}t_{ab})\tau_{bd}
- (\tau_{ac}t_{cd}-\tau_{cd}t_{ac})\tau_{bc}]\left[
X^a_{\a0}X^b_{\b0}X^c_{0\a}X^d_{0\b}-
X^a_{\a0}X^b_{\b0}X^c_{0\b}X^d_{0\a}
\right]\\
 +  \sum_{abcd, \a\b\s}(\tau_{ab}t_{bc}-\tau_{bc}t_{ab})\tau_{cd}\left[ X^a_{\a0}D^b_{\s\a}D^c_{\b\s}X^d_{0\b}
-X^d_{\b0}D^c_{\s\b}D^b_{\a\s}X^a_{0\a}\right],
\end{multline}
\end{widetext}
where we have defined (see also appendix \ref{app:Hubbard}) the bosonic operator $D^i_{\a\b}\equiv(X^i_{\a\b} +\d_{\a\b}X^i_{00})$. All the other operators in the expression are fermionic.

In the following analysis we will show that the average value of this expression and of the other parts of the double commutator is $\sim\d$ at the leading order.

Indeed, after having brought together and collapsed all the operators that refer to the same site into one ($X^a_{\a0}X^a_{0\a}=X^a_{\a\a}$, see appendix \ref{app:Hubbard}), expressions containing  $X^i_{00}$ or fermionic operators vanish exactly as $\d\rightarrow 0$, because of the constraint.

This is an exact statement,  but it can also be seen more explicitly in approximate schemes like slave-bosons, as we will show here.
Indeed one can evaluate the doping dependence of the average of the different terms in the double commutator by considering that each fermionic operator on a different site carries a renormalization factor proportional to $b\sim\sqrt{\d}$, while an operator $X^i_{00}$ carries a factor $|b^2|\sim\d$:
\be\label{eq:HubbardToFermion}
X^i_{\a\b}\! = \!f^\+_{i\a}f_{i\b},\;
X^i_{\a0}\! = \!f^\+_{i\a} b_i,\;
X^i_{0\a}\! = \!b^\+_i f_{i\a},\;
X^i_{00}\! = \!b^\+_i b_i.
\ee

One can then evaluate, using these relations, the explicit dependence in doping of each term of the sum corresponding to the r.h.s. of  (\ref{eq:sumrule_omega}) with the prescription of collapsing first the operators living on same sites into one.

We will see that the coefficients of all terms containing no $X^i_{00}$ and less than two fermionic operators on different sites (that we call "dangerous terms") vanish and thus the dependence is $\propto \d$ at the leading order.

Let's analyse then $[ [\tau^{xy}, H_t], \tau^{xy}]]$.

For the first of the two four-fermion contributions in (\ref{eq:HtTau}) these "dangerous terms" occur when in the sum $a=c$ and $b=d$.
For example the first product in this case contributes to the sum (over $a\neq b,\a\neq\b$) with terms like:
\be\label{eq:simplify}
X^a_{\a0}X^b_{\b0}X^a_{0\a}X^b_{0\b}=-X^a_{\a\a}X^b_{\b\b}=-f^\+_{a\a} f_{a\a} f^\+_{b\b} f_{b\b}
\ee

since in this case one can anticommute the Hubbard operators in order to collapse the ones that refer to the same sites ($X^a_{\a0}X^a_{0\a}=X^a_{\a\a}$, see appendix \ref{app:Hubbard}) and then evaluate the averages using the slave bosons operatorial equivalences(\ref{eq:HubbardToFermion}).

This term is "dangerous" since its average is not obviously proportional to doping. But it doesn't actually contribute to the sum since its coefficient in this case vanishes. Indeed for $a=c$ and $b=d$ its coefficient becomes
$(\tau_{ab}t_{ba}-\tau_{ba}t_{ab})\tau_{bb} - (\tau_{aa}t_{ab}-\tau_{ab}t_{aa})\tau_{ba}=0$,
since $\tau_{ii}=t_{ii}=0$.

The second four-fermion product instead is dangerous when $a=d$ and $b=c$ and again its coefficient $(\tau_{ab}t_{bb}-\tau_{bb}t_{ab})\tau_{ba} - (\tau_{ab}t_{ba}-\tau_{ba}t_{ab})\tau_{bb}=0$.

The two-fermion terms are dangerous when $a=d$ and the coefficient becomes $(\tau_{ab}t_{bc}-\tau_{bc}t_{ab})\tau_{ca}$.  For this to be nonzero $a$ and $c$ have to be next-nearest neighbours. Then it is easy to see that on the cubic lattice (and within the ranges previously defined for the matrices $t_{ij}$ and $\tau_{ij}$) it is impossible to chose the site $b$ such that the coefficient is nonzero.

Thus we have shown that all contributions of $[ [\tau^{xy}, H_t], \tau^{xy}]]$ with a less than linear dependence in doping vanish.

For the magnetic part of the hamiltonian one has:
\begin{widetext}
\begin{multline}\label{eq:HJTau}
\left [ [\tau^{xy}, H_J], \tau^{xy} ]\right]
=\sum_{abcd,\a\neq\b} 4\tau_{ab}\tau_{cd}(J_{bc}-J_{ac}-J_{bd}+J_{ad}) (X^a_{\a0}X^b_{0\b}X^c_{\b0}X^d_{0\a}-X^a_{\a0}X^b_{0\a}X^c_{\b0}X^d_{0\b}) \\
+ \sum_{abcd,\s \a\neq \b}2\tau_{a b}\tau_{bd}(J_{bc}-J_{ac}) \left[ X^a_{\a0} D^b_{\s\b}X^d_{0\s}X^c_{\b\a}
+  X^c_{\a\b}X^a_{\b 0}D^b_{\s\a}X^d_{0\s}
+ X^d_{\s 0} D^b_{\a\s}X^a_{0\b}X^c_{\b\a} \right.\\
\left. +  X^c_{\a\b}X^d_{\s 0}D^b_{\b\s}X^a_{0\a}
 -  X^a_{\a 0} D^b_{\s\a}X^d_{0\s}X^c_{\b\b}
- X^c_{\b\b} X^a_{\a 0}D^b_{\s\a}X^d_{0\s}
  -  X^d_{\s 0}D^b_{\a\s}X^a_{0\a}X^c_{\b\b}
-  X^c_{\b\b}X^d_{\s 0}D^b_{\a\s}X^a_{0\a} \right].
\end{multline}
\end{widetext}
 The first of the four-fermion terms is never dangerous, since even when $a=d$ and $b=c$, the product $X^b_{0\b}X^c_{\b0}=X^b_{00}$ and thus its average is proportional to doping.

 The second of the four-fermion terms is dangerous when $a=b$ and $c=d$. But in these configurations the coefficient becomes $ 2\tau_{aa}\tau_{cc}(J_{ac}-J_{ac}-J_{ac}+J_{ac})=0$.

 The analysis of the two-fermion terms is a little more involved. They are dangerous when $a=d$. The product in this case becomes (in $D^i_{\a\b}\equiv(X^i_{\a\b} +\d_{\a\b}X^i_{00})$ we can drop the $X_{00}$ operators\footnote{When averaged, an $X_{00}$ operator leads a renormalization factor $\propto \d$, unless if it shares the site index with other operators in the product, in which case it has to be simplified as in eq. (\ref{eq:simplify}). In the expression under examination it is never the case, since when  $a,d = b$ the coefficient vanishes given the presence of $\tau_{ab}\tau_{bd}$}):
 \bea
 &&X^a_{\a\s}X^b_{\s\b}X^c_{\b\a} + X^c_{\a\b}X^b_{\s\a}X^a_{\b\s} \nonumber\\
 &+& X^a_{\s\b}X^b_{\a\s}X^c_{\b\a} + X^c_{\a\b}X^b_{\b\s}X^a_{\s\a}\nonumber\\
 &-&X^a_{\a\s}X^b_{\s\a}X^c_{\b\b} - X^c_{\b\b}X^b_{\s\a}X^a_{\b\s} \nonumber\\
 &-&X^a_{\s\a}X^b_{\a\s}X^c_{\b\b} - X^c_{\b\b}X^b_{\a\s}X^a_{\s\a}\nonumber.
  \eea
 We used the fact that operators at sites $a$ and $d$ always commute with operators at site $b$ since the coefficient vanishes when $a=b$ or $d=b$.

 We then have to exchange the indices $a$ and $b$ in the III, IV, VII and VIII term, in order to match the spin indices with the remaining four terms (the operators in $a$ and $b$ commute). We obtain:

  \bea
 &&X^a_{\a\s}X^b_{\s\b}X^c_{\b\a} + X^c_{\a\b}X^b_{\s\a}X^a_{\b\s} \nonumber\\
 &-& X^a_{\a\s}X^b_{\s\b}X^c_{\b\a} - X^c_{\a\b}X^b_{\s\a}X^a_{\b\s}\nonumber\\
 &-&X^a_{\a\s}X^b_{\s\a}X^c_{\b\b} - X^c_{\b\b}X^b_{\s\a}X^a_{\b\s} \nonumber\\
 &+&X^a_{\a\s}X^b_{\s\a}X^c_{\b\b} + X^c_{\b\b}X^b_{\s\a}X^a_{\b\s} \nonumber =0,
  \eea

  since the renaming of $a$ and $b$ changes the sign of the coefficient $\tau_{a b}\tau_{bd}(J_{bc}-J_{ac})$. Thus the terms cancel two by two (I and III, II and IV, V and VII, VI and VIII).

We can then conclude that all the terms that contribute to $\langle [ [\tau^{xy}, H], \tau^{xy}]]\rangle$ are at least
$\propto \d$ and that the sum rule
integral (\ref{eq:sumrule_omega}) is proportional to doping.
Note that this is true {\it irrespectively of the nature of the possible long-range
ordering} (e.g. both in the normal and superconducting states).

\section{DMFT calculation in the Hubbard model}\label{sec:DMFT}

In this section we report the explicit calculation of the Raman response function for the doped Hubbard model,
\be
H=\sum_{ij,\s} t_{ij} (c^\+_{i\s}c_{j\s} + h.c.) + U\sum_i n_{i\up}n_{i\down},
\ee
at strong and intermediate coupling within Dynamical Mean-Field Theory\cite{Georges_RMP_DMFT} (DMFT).
One of our main motivations for studying the Hubbard model is to investigate the {\it restricted}
sum-rule, up to a frequency cutoff $\Omega_c$, and study the dependence on the cutoff frequency.

Single-site DMFT freezes spatial fluctuations while fully retaining the local dynamics. It thus allows to calculate dynamical response functions in the local self-energy approximation.
\footnote{We are interested here in the Raman response of a bidimensional cubic  lattice, so in the calculation of the response function we will make use of a form factor obtained by integrating functions of the cubic dispersion (\ref{eq:disp}) over the bidimensional Brillouin zone.
Hence DMFT here has to be thought as a local self-energy approximation of the finite dimensional lattice, rather than the solution of the $\infty$-dimensional counterpart of the model under investigation.
In this light the choice of a semicircular DOS (see the footnote on DMFT technical data) for the impurity model (that facilitates the numerical solution) is not a severe inconsistency with the cubic dispersion used in the response function.}

Since the Raman vertex is odd under e.g. $k_x\rightarrow -k_x$,
vertex corrections vanish in the local self-energy
approximation\cite{Georges_RMP_DMFT} and the stress-stress response function eq. (\ref{Kubo_susc}) at ${\bf q}=0$ reduces to the simple bubble-diagram of dressed propagators, that reads (in the imaginary time formalism):
\be
\chi (i\Omega) =\frac{1}{\b}\sum_{\vk,i\nu} (\frac{\partial^2 \epsk}{\partial k_x\partial k_y})^2 G(\vk,i\nu)G(\vk,i\nu+i\Omega),
\ee
and analytically continued,
\be\label{eq:chisecond}
\chi\" (\Omega) = \# \int \! d\eps V(\eps) \# \int \! d\omega A(\eps,\omega) A(\eps,\omega+\Omega)\left[ f(\omega) \! - \! f(\omega+\Omega) \right],
\ee
where
\bal
V(\eps)&=\sum_\vk(\frac{\partial^2 \epsk}{\partial k_x\partial k_y})^2\delta(\eps-\epsk)\nonumber\\
&=(4t^\')^2\sum_\vk \sin^2{k_x}\sin^2{k_y}\delta(\eps-\epsk)
\end{align}
is the appropriate form factor obtained by summing over k the product of two $B_{2g}$ geometry vertices eq. (\ref{eq:B2gVertex}). It can be calculated once and for all and for the dispersion (\ref{eq:disp}) is a smooth function. $A(\eps,\omega)$ is the spectral function\bal
A(\vk,\omega)&=-\frac{1}{\pi} Im G(\vk,\omega)\nonumber \\
&=-\frac{1}{\pi}\frac{Im\Sigma(\o)}{(\o+\mu-\epsk-Re \Sigma(\o))^2+(Im\Sigma(\o))^2} ,
\end{align}
for $\epsk=\eps$ and is easily accessible in single site DMFT\footnote{The Anderson impurity model associated by DMFT to the Hubbard model has been solved by means of the Exact Diagonalization (ED) of the Hamiltonian.
The impurity model Hamiltonian has been limited to 8 sites (inpurity + 7 sites in the effective bath), and it has been diagonalized at zero temperature by means of the Lanczos algorhythm. The projection of the Anderson bath on the limited subspace is obtained by means of a best fit on a grid of frequencies on the imaginary axis, corresponding to a fictive temperature of $\beta=500$, and slightly weighted towards low frequencies.
The density of states (DOS)  entering the bath selfconsistency condition is semicircular, of bandwidth W.
The self-energy obtained from the impurity model with this choice of the density of states equals the self-energy of an infinite dimensional Bethe lattice.}\label{note:DMFT}, through the knowledge of the local self-energy\footnote{The real-axis self-energies are calculated for $\omega=\omega+i\Gamma_{br}$, with $\Gamma_{br}=0.1$ in order to smoothen out the spikes due to the discretization of the bath.}.

Hence in this approximation the non-resonant Raman response is readily calculated by convoluting two single particle spectral functions with the appropriate kernel $V(\eps)$.

We have solved the Hubbard model with DMFT at zero temperature and finite doping for $U/W=2.5$ (W is the bandwidth), that can be seen as a rather strong coupling, and for an intermediate coupling, $U/W=1.35$.

\begin{figure}[htbp]
\begin{center}
\includegraphics[width=9cm]{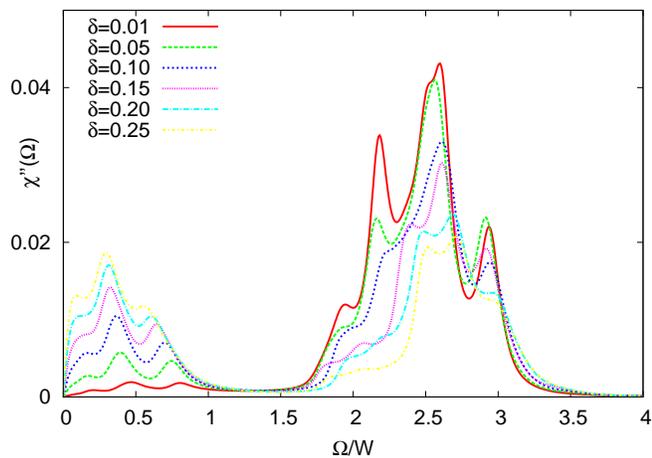}
\end{center}
\caption{(Color online) Evolution of the $B_{2g}$ Raman response function for $U/W=2.5$ with doping ranging from $\d=0.01$ to $\d=0.25$. The low energy feature increase in weight for increasing doping, the high energy one decreases.}
\label{fig:Raman_resp_doping}
\end{figure}
Let's first focus on the strong coupling case.
Indeed this correspond to a doped Mott insulator, the critical U at half-filling for the chosen density of states (see below) being $U_c/W\simeq1.5$, and at $U/W=3$ the Hubbard bands being split apart by a gap $\Delta \simeq 1.5W$
\footnote{The spectral function $A(\eps,\omega)$ of the doped Hubbard model has a well known three-peak structure\cite{Georges_RMP_DMFT} formed by a resonance peak close to the chemical potential (in the metallic phase) and two features lying, at an energy $\sim U$ from one another, in the "Hubbard bands".
This two energy intervals are of width W and are separated by a gap $\Delta \sim U-W$. In the doped phase the chemical potential lies within one of them.}.

In figure \ref{fig:Raman_resp_doping} we show how the response function calculated in DMFT shows two principal features. These are expected when $\Delta>W$ from the general features of $A(\eps,\omega)$ and the smooth form of $V(\eps)$.

At zero temperature the Raman response formula simplifies further in:
\be\label{eq:chisecond_simple}
\chi\" (\Omega) =\int d\eps V(\eps)\int^0_{-\Omega} d\omega A(\eps,\omega) A(\eps,\omega+\Omega)
\ee
The integral will be nonzero when, for some $\eps$, the two spectral functions shifted by $\Omega$ from one another have an overlap in the region $[-\Omega,0]$, i.e. between the two quasiparticle resonances.

Hence at low $\Omega$, $\chi\"$ will show contributions from the overlap of the two functions within the same Hubbard band, the one in which the quasiparticle peak lies.

At higher $\Omega \gtrsim W$, since the width of the Hubbard band is $\sim W$, if the gap is bigger than the bandwidth ($\Delta>W$), the two spectral functions will have negligible overlap. Indeed the Hubbard band of one will fall onto the gap of the other,  and the response will be nearly zero.
Then for even bigger $\Omega$ the two spectral functions will be shifted enough so to have the upper Hubbard band of one overlapping the lower Hubbard band of the other,  and a finite response is found, thus generating a second, well separated feature at these higher frequencies.

If instead $\Delta<W$, this separation of the Response will in general not occur.

The evolution of this response function with doping is such that the lower energy feature grows with increasing doping, while the higher energy one decreases. This is expected, since upon doping the system dilutes and becomes less and less correlated thus causing a transfer of spectral weight back from high to low frequencies.

Physically the separation of the response function into two features means that the Coulomb repulsion is large enough to separate in energy the particle-hole excitations happening in the Hubbard band of empty sites from the ones involving doubly occupied sites.

In the strong coupling limit and at low doping this model can be mapped on a t-t'-J model and this allows us to identify the low-frequency response with that of the t-J model, so that we can use it to confirm our analysis on the sum-rule integral.

We thus consider the response under a cutoff $\Omega_c\sim W$, that includes only the processes of the low-energy feature of the Raman Response function, namely
\be
W(\Omega_c) \equiv \!\! \int_0^{\Omega_c}\@ d \Omega \Omega \chi"_{Hub}(\Omega)
\simeq\!\!\int_0^\infty\@ d \Omega \; \Omega \; \chi"_{tJ}(\Omega),\: \mbox{if}\; \Omega_c\gtrsim W.
\ee

\begin{figure}[htbp]
\begin{center}
\includegraphics[width=9cm]{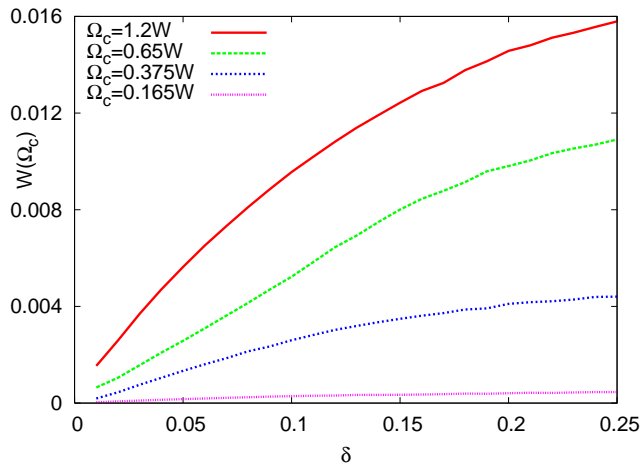}
\end{center}
\caption{(Color online) Integrated spectral weight up to a cutoff $\Omega_c=1.2W$, that fully includes the lower energy feature) of the first moment of the Raman response function, plotted versus doping. As predicted for the t-J model $W(\d)\propto \d$. Integral up to lower cutoffs are also plotted.}
\label{fig:Int_Raman}
\end{figure}

In fig. \ref{fig:Int_Raman} we plot the calculated $W(\Omega_c)$ as a function of different cutoff frequencies $\Omega_c$. The highest value, $\Omega_c=1.2W$ fully includes the low-energy feature of the Raman Response in our case, thus reproducing the t-t'-J-model sumrule.

It appears that the linearity in doping of $W(\Omega_c\simeq W)$ holds well up to dopings of about 10\%, then higher order terms start contributing considerably.

By reducing the cutoff frequency one can see that the integral is still linear in doping. This is somehow expected, since the whole lower feature scales with doping, and vanishes as the insulating phase is reached for $\d=0$.
The linearity region actually increases and extends up to 15\% doping (i.e. the whole underdoped region in cuprates).

The smaller extent of the linear region for the higher cutoffs is probably due to the proximity of the high-energy feature in the response function, that would indeed be completely absent if we were to take the actual $U/W\rightarrow \infty$ limit.

But in fact considering a cutoff lower than $\Omega_c\simeq W$ is actually directly relevant to the
analysis of experimental data on cuprates, as we will see in section \ref{sec:exp}.

Let's now consider the intermediate coupling case.

\begin{figure}[htbp]
\begin{center}
\includegraphics[width=9cm]{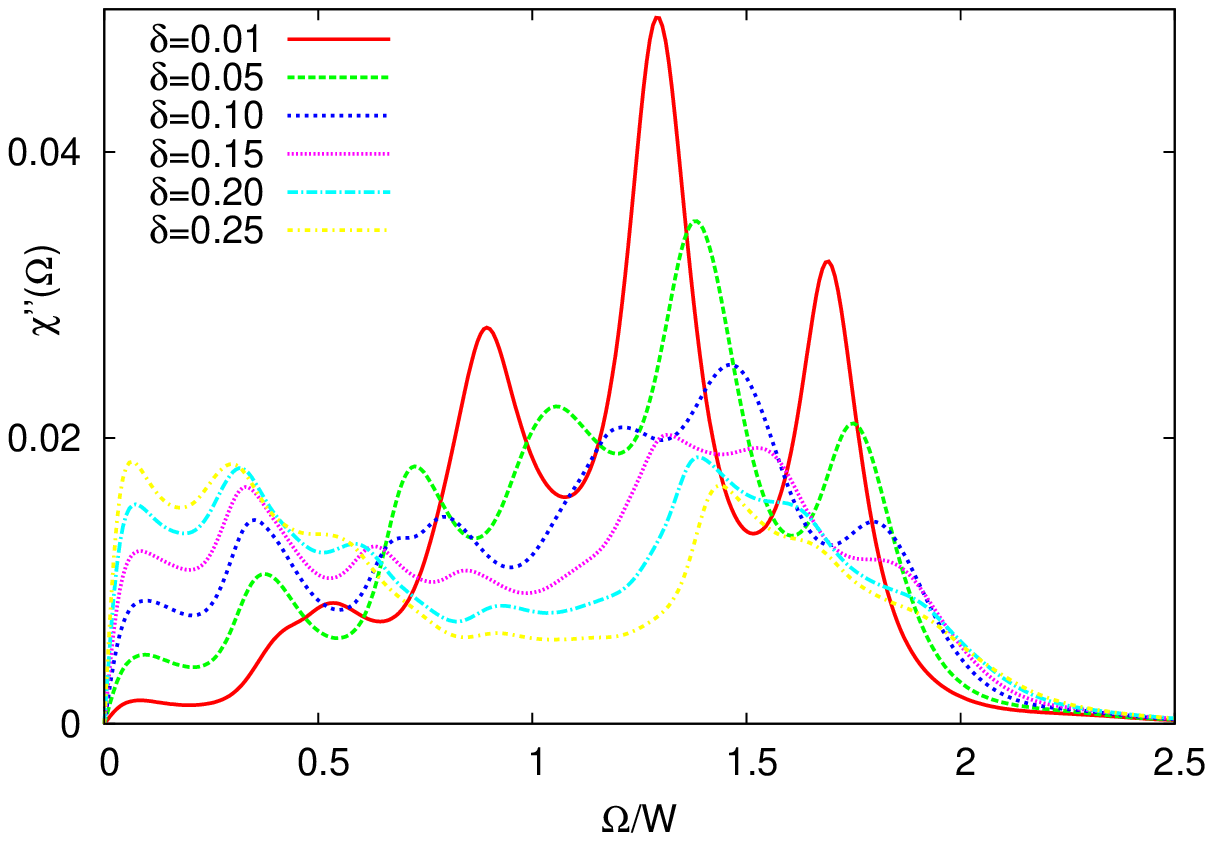}
\includegraphics[width=9cm]{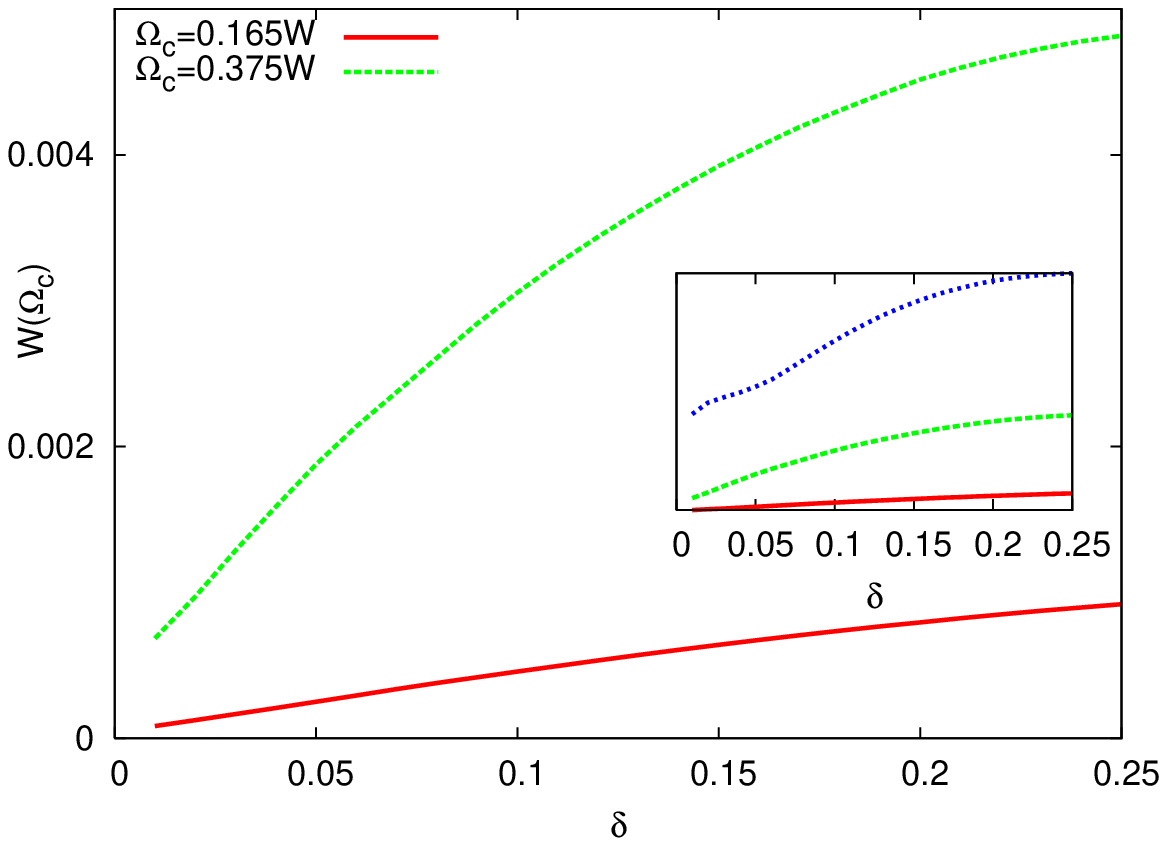}
\end{center}
\caption{(Color online) {\bf Upper panel}: non-resonant $B_{2g}$ Raman response function for $U/W=1.35<U_{c2}$. Different curves are, from bottom to top, for increasing dopings $\d=0.01\div0.25$.
The two separate features present at large U/D have now partially merged, but the curves still scale with doping at low frequency. {\bf Lower Panel}: sum-rule integral for cutoffs $\Omega_c=0.165W$ (lower curve) and $\Omega_c=0.375W$ (upper curve). {\bf Inset}: sum-rule integral for all the chosen cutoffs in Fig. \ref{fig:Int_Raman}: from bottom to top $\Omega_c/W=0.165, 0.375, 0.65, 1.2$. An intercept is clearly non negligible for $\Omega_c=0.65W$, the behaviour is no more linear for $\Omega_c=1.2W$ of the order of the bandwidth.}
\label{fig:chi_Raman_U2.7}
\end{figure}

We studied the model at $U=1.35W=0.9U_{c2}$ and obtained the results plotted in the upper panel of fig. \ref{fig:chi_Raman_U2.7}.
One imediately sees that the two features of the strong coupling case have now merged as the separation of the two Hubbard bands is less than $W$.

Nevertheless, the low energy part of $\chi^\"(\Omega)$ still scales with doping.

In the lower panel of fig. \ref{fig:chi_Raman_U2.7} we plot the value of the integral $W(\Omega_c)$ for small cutoffs as a function of doping. Although the intercept is indeed nonzero, owing to the fact that a quasiparticle resonance is still present at half-filling in this case (that testifies the metallic behavior at half-filling that one obtains when $U$ is smaller than the Mott transition critical value $U_{c2}$), this intercept is practically negligible for the lower cutoffs, given the very little weight of the quasiparticle resonance very close to $U_{c2}$. Thus the behaviour of the sum-rule integral for small cutoffs is still $\simeq \delta$ in the underdoped region for $U\simeq 0.9U_{c2}$.

Indeed by raising the cutoff to values closer to the bandwidth one eventually loses the linear behaviour, owing to the presence of the low-frequency tail of the spectrum of transitions between the two Hubbard bands.

\section{Connection with experiments: cutoff frequency and coupling strength}\label{sec:exp}

\begin{figure}[htbp]
\begin{center}
\includegraphics[width=9cm,height=13cm]{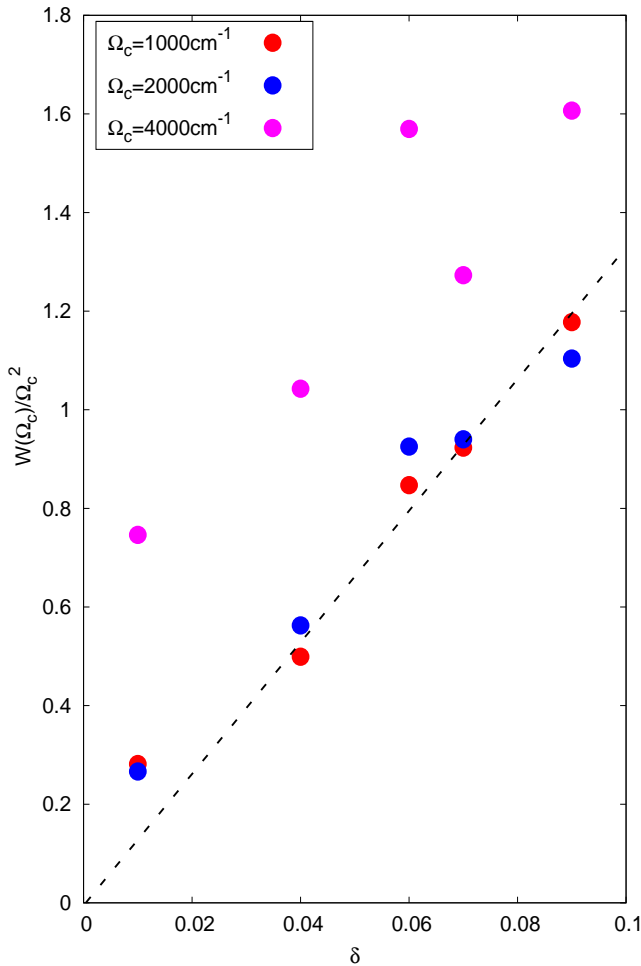}
\end{center}
\caption{(Color online) Sumrule integral $W(\Omega_c)$ calculated from the experimental
data on $(Y_{1-y}Ca_y)Ba_2Cu_3O_{6+x}$ in the normal phase, taken from
refs \cite{Opel_Raman}and \cite{Tassini_Hackl_2007} where absolute intensities
of Raman scattering have been measured. The plots are rescaled by normalizing to
the square of the cutoff frequency. The sum-rule linearity holds clearly up to $\d \sim 0.1$ for
both lower cutoffs. The line is a least square fit of the $\Omega_c=1000 cm^{-1}$ datapoints,
discarding the $\d=0.01$ point that corresponds to the antiferromagnetic phase.
The curves for these lower cutoffs extrapolate linearly to zero or to a very small intercept.
The curve obtained for $\Omega_c=4000cm^{-1}\sim 0.5 eV$ instead doesn't seem to extrapolate to zero.
As discussed in the text, this may indicate either that cuprates are in an intermediate
coupling regime\cite{Comanac}, or that additional magnetic contributions (e.g. two magnons) contribute
to the Raman signal in that frequency range.}
\label{fig:Data_Hackl_linearity}
\end{figure}

Some experimental setups are believed to give measures that are accurate enough to
extract reliably absolute scattering intensities.
We have compiled and anayzed experimental data from two such references\cite{Opel_Raman}\cite{Tassini_Hackl_2007}, which
are shown in figure \ref{fig:Data_Hackl_linearity}.
The linear dependence of the absolute intensities in the underdoped region (at least up to $\sim 10 \% $ doping) is clearly visible for small cutoffs, up roughly to $2000 cm^{-1}\sim 0.25 eV$.

Hence if the experimental cutoff is as low as $\Omega_c\lesssim 2000 cm^{-1}\sim 0.25eV$ Raman spectra can indeed be normalized using a linear scaling in doping, as was done in Ref. \cite{LeTaconNaturePhysics}.

By raising the cutoff around $4000cm^{-1}\sim 0.5eV$ one finds a non vanishing intercept as is visible in fig.\ref{fig:Data_Hackl_linearity}.

It is also very interesting to notice the nonvanishing intercept found for higher cutoffs.
This is consistent with the very recent analogous findings of Comanac et al. in Ref. \cite{Comanac},
where by analyzing optical conductivities of a wide range of cuprates the mid-infrared spectral weight ($\Omega_c=0.8eV$) is found to be non vanishing despite a vanishing Drude spectral weight ($\Omega_c=0.2eV$).
To explain these experiments these authors propose to model the optical conductivity of cuprates with a Hubbard model whose interaction strength is less than the single-site DMFT  $U_{c2}\simeq1.5W$. This is the value at which a pure Mott transition takes place, without additional effects due to magnetic correlations.

Indeed the calculation we performed in section \ref{sec:DMFT}, at $U=1.35W<U_{c2}$, reproduces the behavior found in the experiments we reviewed. As is clear in the lower panel of fig. \ref{fig:chi_Raman_U2.7}
while for intermediate cutoffs an intercept is sizeable, for the lower cutoffs it is instead very small.
Moreover the antiferromagnetic correlations neglected in this single-site DMFT approach are likely to depress the low-frequency data, thus further reducing the intercept.

%
There are two possible interpretations of this observation of a finite intercept for the higher
values of the cutoff (keeping in mind however that the Raman data used here are taken at slightly different
temperatures, and that more data are needed).

The first one is that cuprates are actually in the regime of intermediate correlation
strength, as recently advocated by Comanac et al.\cite{Comanac} on the basis of a similar observation
from optics.
%
%
Another possible interpretation of the nonvanishing intercept found for $\Omega_c=4000 cm^{-1}$
are additional contributions to the Raman scattering of spin origin, such as two magnon processes.
To settle this issue clearly further work is needed.

\section{The $B_{2g}$ sum-rule in the superconducting phase of the $t-t^\'-J$ model: Slave bosons}\label{sec:KLSB}

In this paragraph we calculate explicitly the right-hand side of eq. (\ref{eq:sumrule_omega}), by averaging the calculated expressions for the commutators, eq. (\ref{eq:HtTau}) and (\ref{eq:HJTau}), in the Slave-bosons mean-field formalism of Ref. \cite{kotliar_liu_RVB} used in the early studies of High-Tc's, that is the simplest approximation that allows to access the d-wave superconducting phase of the t-t'-J model, beside the normal phase.

We recall here briefly this approach.
The hamiltonian of the t-t'-J model, eq. (\ref{eq:tb_Ham}), lives in the restricted Hilbert space in which all states with a doubly occupied site are projected out. This can be expressed by means of a mixed fermion-boson hamiltonian:
\begin{align}\label{eq:fermion-boson}
H&=\sum_{i,j,\s}(t_{ij} f^\+_{i\s}b_ib^\+_jf_{j\s} + H.c.) -\mu\sum_{i\s}f^\+_{i\s}f_{i\s}\nonumber\\
&+\sum_{ij}4J_{ij}[(S_i\cdot S_j)-\frac{1}{4}(1- b^\+_ib_i)( 1-b^\+_jb_j)]\nonumber\\
&+\sum_{i}\lambda_i\left( \sum_\s f^\+_{i\s}f_{i\s}+b^\+_ib_i-1 \right) ,
\end{align}
in which the boson operators $b^\+_i$ ($b_i$) represent the creation (destruction) of a hole on site $i$, and thus carry the charge degrees of freedom, while the pseudofermions $f^\+_{i\s}$ ($f_{i\s}$) carry the spin information. This hamiltonian lives in the \emph{enlarged} Hilbert space represented by all the possible states of the bosonic and pseudofermionic fields. To complete the mapping a constraint has to be introduced, in order to restrict again this Hilbert space to the physical one, by excluding all the non-physical states.
In doing this, one can also implement the no-double-occupancy condition, using namely:
\be
\sum_\s f^\+_{i\s}f_{i\s} + b^\+_i b_i=1
\ee
 This is expressed in eq. (\ref{eq:fermion-boson}) by means of the Lagrange multipliers $\lambda_i$, that enforce the condition on every site.

One then performs an Hartree-Fock-Bogoliubov decoupling and a static mean field (such that $\lambda_i=\lambda$ and the constraint is satisfied on the average, i.e. $\langle b^\+_i b_i\rangle=1-\sum_\s \langle f^\+_{i\s}f_{i\s}\rangle=\d$) on this Hamiltonian by introducing the following mean-field parameters:
\bal\label{eq:MFParam}
K&=3J\langle \sum_\s f^\+_{i\s} f_{i+x,\s}\rangle/2\nonumber\\
\Delta_x&=-\Delta_y\equiv \Delta=3J\langle f_{i\uparrow} f_{i+x\downarrow}-f_{i\downarrow} f_{i+x\uparrow}\rangle/2\nonumber\\
\langle b^\+_i b_j\rangle &=|\langle b_i\rangle|^2=1-\sum_\s \langle f^\+_{i\s}f_{i\s}\rangle=\d
\end{align}

We choose the phase in which $K$ is isotropic (i.e. $K_x=K_y\equiv K$) while the superconducting order parameter has a d-wave symmetry ($\Delta_x=-\Delta_y$). This has been shown to be the actual ground state at finite doping between all the possible symmetries in this kind of mean-field\cite{kotliar_liu_RVB}.

The obtained mean-field Hamiltonian reads:
\be\label{eq:H_MF}
H_{MF}=\sum_{\vk\s}\epsk f^\+_{k\s}f_{k\s}+(\sum_{\vk\s}\Delta_\vk f^\+_{\vk\up}f^\+_{-\vk\down}+H.c.),
\ee
where

\be\label{eq:renormdisp}
\epsk=-2(\d t-K)(\cos k_x+\cos k_y) + 4\d t^\'\cos k_x \cos k_y
\ee
\be\label{eq:Deltak}
\Delta_\vk=2\Delta(\cos k_x-\cos k_y )
\ee

Eqq. (\ref{eq:MFParam}), (\ref{eq:H_MF}), (\ref{eq:renormdisp}) and (\ref{eq:Deltak}) are the mean-field equations, that have to be iterated until a self-consistent solution is found.

The hamiltonian (\ref{eq:H_MF}) is of the standard Bogoliubov quadratic form, and thus it is trivially solvable (by means of a Boboliubov substitution)
One can then calculate any fermionic average by means of a Wick factorization in terms of the quadratic averages
\bal\label{eq:quadr_av}
\langle f^\+_{i\s} f_{j\s} \rangle &=\sum_\vk \cos (\vk\cdot \vr_{ij})\langle f^\+_{\vk\s} f_{\vk\s} \rangle, \nonumber\\
\langle f_{i\s} f_{j\bar \s} \rangle &=\sum_\vk \cos (\vk\cdot \vr_{ij})\langle f_{\vk\s} f_{-\vk\bar\s} \rangle.
\end{align}

\begin{figure}[htbp]
\begin{center}
\includegraphics[width=9cm]{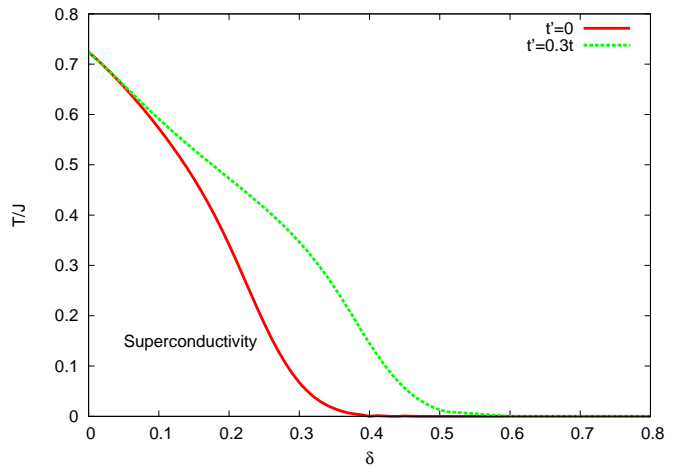}
\end{center}
\caption{(Color online) Doping-temperature phase diagram of the  $t-t'-J$ model within the Kotliar-Liu slave-bosons mean-field approximation for $J=0.1t$ and $t'=0$ and $t'=0.3t$. The actual superconducting region is below a condensation temperature of the bosons, where phase coherence sets in. This temperature is an increasing function of doping, so that the superconducting region is dome-shaped\cite{kotliar_liu_RVB}.}
\label{fig:phDiag_KLSBMF}
\end{figure}

This mean-field hamiltonian of the t-t'-J model has the phase diagram shown in Fig.\ref{fig:phDiag_KLSBMF} for $t=10J$ and for $t^\'=0$ (Kotliar-Liu result) and $t^\'=0.3t$ (the value generally used for High-Tc superconductors like BiSCCO).

For each doping there are two phases, a low temperature (superconducting) phase with $\Delta\neq 0$ and a high-temperature (normal) phase with $\Delta=0$. The transition temperature is denoted $T_{RVB}$ and is a decreasing function of doping. The bosons are treated here as fully condensed, because $\langle b^\+_i b_i\rangle=|\langle b_i\rangle|^2=\d$ at all temperatures.

If we were to allow bosons decondensation
(which requires to include the link variational parameter $\langle b^+_ib_j\rangle$ as well),
another crossover temperature (zero at half-filling and rapidly increasing with doping) would delimit a low temperature phase with phase coherence from the disordered phase of incoherent pairs. The actual superconducting region has a dome-like structure centered around a region of "optimal" doping where the two characteristic temperatures cross.

In this formalism it is then possible to calculate the r.h.s. of eq. (\ref{eq:sumrule_omega}) by taking each Hubbard operator product in the two parts of the commutators, eq. (\ref{eq:HtTau}) and (\ref{eq:HJTau}) and obtain the expression in terms of the $f$'s and the $b$'s using eqq. (\ref{eq:HubbardToFermion}). Then one performs the Wick factorization of each product (i.e. following the example of eq. (\ref{eq:simplify}) one obtains $\langle f^\+_{a\a} f_{a\a} f^\+_{b\s} f_{b\s}\rangle=\langle f^\+_{a\a} f_{a\a}\rangle\langle f^\+_{b\s} f_{b\s}\rangle-\langle f^\+_{a\a} f_{b\s}\rangle\langle f^\+_{b\s} f_{a\a}\rangle-\langle f^\+_{a\a} f^\+_{b\s} \rangle\langle f^\+_{a\a} f_{b\s}\rangle$) and by means of the quadratic averages eq. (\ref{eq:quadr_av}) calculated with $H_{MF}$ within the converged solution one obtains the final result.

\begin{figure}[htbp]
\begin{center}
\includegraphics[width=9cm]{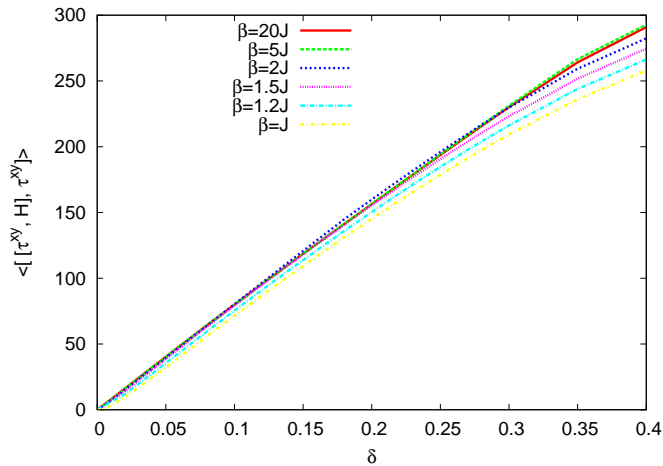}
\end{center}
\caption{(Color online) Right hand side of the sumrule evaluated in the  $t-t'-J$ model within the Kotliar-Liu slave-bosons mean-field approximation for different temperatures and parameters $J=0.1t, t'=0.3t$. The bottom curve is calculated for $\beta=J$, entirely in the normal phase, and shows how the linearity holds pretty well also for the normal phase in the whole range of dopings.}\label{fig:SBMFrhs}
\label{fig:KLSB_linear}
\end{figure}

This result is plotted in Fig.\ref{fig:SBMFrhs} as a function of doping at different temperatures.

 The results shows that in the superconducting phase at $T<T_{RVB}$ the sum-rule integral is linear in doping until 30-40\% doping, thus allowing the linear scaling of Raman spectra in doping for the full extent of the superconducting phase of cuprates.

 Still one has to be aware of the many limitations of this simple mean-field.  Beside the cited boson decondensation, that adds a second energy scale to the problem, less relevant to our analysis, the method lacks a second \emph{fermionic} energy scale that is the new and very debated point put forward in \cite{LeTaconNaturePhysics}. Our analysis in terms of the Kotliar-Liu slave-bosons has to be taken thus as a qualitative description of the nodal physics in the superconducting phase.

It is also interesting to plot the temperature dependence of the sumrule integral.

\begin{figure}[htbp]
\begin{center}
\includegraphics[width=9cm]{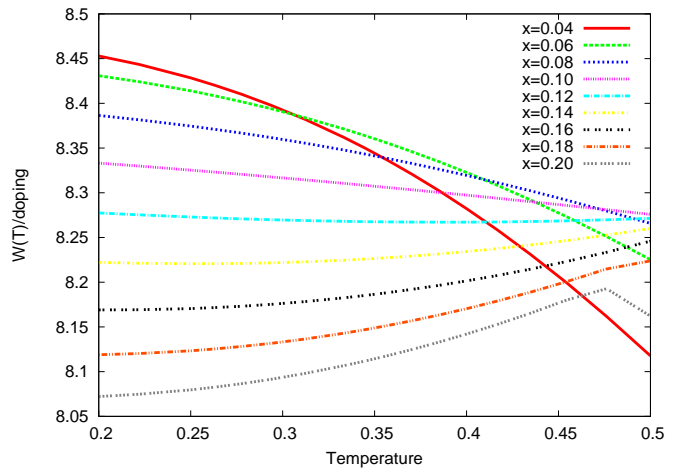}
\end{center}
\caption{(Color online) Temperature dependence of the sumrule in the Slave bosons formalism in the superconducting phase. For increasing temperature the integral decreases at low doping while it increases at higher doping.}
\label{fig:KLSB_temp}
\end{figure}

In order to compare the temperature dependence in the superconductive phase at different dopings, in fig. \ref{fig:KLSB_temp} we plot $\langle \left [ [\tau^{xy}, H], \tau^{xy} \right]\rangle/\d$, owing to the doping linearity. The rescaled curves indeed lie all in the same range of values but show a different behavior in temperature depending on the doping value. While at low doping raising the temperature causes a decrease of the sumrule value, after $\d\sim12\%$ (that could be interpreted as "optimal doping" in this Slave Bosons mean-field ) the value instead increases with increasing temperatures.
Then in general when the system enters the normal phase upon heating the integral value shows a cusp and then drops quickly.

This behaviour has not yet been measured and it is a prediction of this slave boson mean-field treatment that would be very interesting to clarify in further experiments.

\section{Conclusions}

%
We have studied a restricted sum-rule for Raman scattering, which involves the first moment of the
non-resonant $B_{2g}$ response function up to a cutoff frequency:
$\int_0^{\Omega_c} d \Omega \; \Omega \; \chi_{B_{2g}}^\"(\Omega)$.

For the doped t-t'-J model, where we can take the cutoff $\Omega_c=\infty$,
we have shown that this integral is proportional to the doping level.
This supports the experimental data analysis of Ref.~\cite{LeTaconNaturePhysics}.

We have then calculated the Raman response function and studied its doping dependence for
the normal phase of the Hubbard model within DMFT.

We have first studied the system in the strong coupling case $U=2.5W$, that corresponds to a doped Mott insulator.
In this case the low-frequency response of this system can be identified with the one of the t-t'-J model, and we have confirmed its linearity in doping at low doping.

We have then studied the intermediate-coupling case $U=1.35W=0.9U_{c2}$, showing that in this case the sum-rule integral still shows a behavior $\simeq \d$  for very low cutoffs, while the intercept is non-negligible for intermediate cutoffs.

We have also shown a compilation of experimental data that are believed to correctly measure the absolute intensity of Raman scattering on YBCO.
As in our DMFT study at $U=1.35W$, the behaviour is $\simeq \d$ at low cutoffs ($\Omega_c\lesssim 0.125 eV$) but shows a non-negligible intercept already for $\Omega_c\sim 0.25 eV$. This can be interpreted in support of the hypotesis that cuprates are ``intermediately correlated'' as recently put forward in Ref. \cite{Comanac}.

Finally we have studied the superconductive phase of the t-t'-J model using the slave-boson method of Ref. \cite{kotliar_liu_RVB}. We can calculate explicitly the right-hand side of the sum-rule in this approximation and we show its linearity in doping up to dopings of order of 30\%, and a prediction on the temperature dependence.

The combination of those results give a  first indication on
how the strong correlations affect  integrated Raman intensities.
Missing at this point,  a systematic study
of  the effect of  dynamical short range magnetic correlations,
since the slave boson method captures these in a primitive fashion
using static link expectation values. These limitations can
be removed using cluster versions of dynamical mean field theory\cite{BiroliParcolletKotliar_CDMFT} and this  is left for future studies.

\section{Acknowledgements}
LdM thanks A. Camjayi, L. De Leo and M. Capone for useful discussions.
AG and GK acknowledge an earlier collaboration and discussions with A. Sacuto and M. Le Tacon,
as well as discussions with N. Bontemps.
This work was supported by the NSF under Grant No. DMR 0528969, and by the Blaise Pascal Chair.

\appendix
\section{Some algebra of the Hubbard operators used in this paper}\label{app:Hubbard}
Hubbard operators on site i are defined as:
\be
X^i_{\a\b}=|\a\rangle_i\langle\b|_i; \qquad |\a\rangle,|\b\rangle = |0\rangle, |\up\rangle, |\down\rangle
\ee
and hence they are used to enforce the constraint of no onsite double occupancy, since they all project the $|\up\down\rangle$ state to zero.

From their definitions it is easy to show that two Hubbard operators acting on the same site observe the following commutation (anticommutation) relations:
\be\label{eq:comm_onsite}
[X^i_{\a\b}, X^i_{\g\d}]_\pm=\d_{\b\g}X^i_{\a\d}\pm \d_{\a\d}X^i_{\g\b}
\ee

Instead two operators acting on different sites respect the canonical commutation or anticommutation relations depending on them being "fermionic" ($X_{0\s}$ or $X_{\s 0}$ that add or remove a spin-1/2) or "bosonic" ($X_{00},X_{\s\s'}$ that either don't change the onsite spin or they add or remove a spin 1), i.e.
\be
{[X^i_B,X^j_B]}_- \@=0, \;
{[X^i_F,X^j_F]}_+ \@=0, \;
{[X^i_B,X^j_F]}_- \@=0, \:\forall \; i\neq j
\ee
where "B" is for "bosonic" and "F" is for "fermionic". Indeed these operators are not actual bosons or fermions since on-site they respect the (\ref{eq:comm_onsite}).

It is useful in our calculations to introduce the "bosonic" operator $D^i_{\a\b}$ corresponding to the particular case of (\ref{eq:comm_onsite}) for the onsite anticommutation of two fermionic operators:
\be
[X^i_{\a0}, X^i_{0\b}]_+= X^i_{\a\b}+ \d_{\a\b}X^i_{00} \equiv D^i_{\a\b}
\ee

and also its on-site commutation relations with the others Hubbard operators:

\be
{[D^i_{\s\s'}, X^i_{\a\b}]}_-= \d_{\s'\a}X^i_{\s\b}- \d_{\s\b}X^i_{\a\s'} +\d_{\s\s'}(\d_{\a 0}X^i_{0\b}-\d_{\b 0}X^i_{\a 0}).
\ee

\bibliographystyle{apsrev}

\end{document}